# A Computational Study of Bubble Formation from an Orifice Submerged in Liquid with Constant Gas Flow

James Q. Feng

OXCBO Research, Maple Grove, Minnesota, USA

james.q.feng@gmail.com

**Abstract**

The process of bubble formation from an orifice submerged in liquid with constant gas flow is studied by numerical simulations using an OpenFOAM® volume-of-fluid solver named interIsoFoam. The computed results show that the detached bubble size tends to increase with the gas flow rate, orifice size, surface tension, liquid contact angle, etc. in qualitative agreement with most previous authors. For a given orifice size and liquid properties, there exists a critical gas flow rate above which detached bubbles will combine via coalescence known as bubble pairing. At low gas flow rates, the volume of detached bubbles in the quasi-static regime is shown to depend linearly on the gas flow rate, consistent with a physical mechanistic analysis but not recognized by previous authors. The detached bubble size seems insensitive to the contact angle when the liquid adequately wets the orifice wall, but can increase substantially if the contact angle is increased beyond a critical value resulting in contact line motion on the horizontal outside wall of orifice. The value of such a critical contact angle is found to increase with the orifice size and decrease with the gas flow rate. Such revelations would logically suggest that reducing orifice size for generating smaller bubbles could be more challenging for sub-millimeter orifices with constant gas flow.

## 1 Introduction

Flowing a gas through an upward-facing orifice into a liquid tank usually generates a sequence of bubbles, which has become essential in a wide range of phase contacting applications as seen in bubble columns, sparger reactors, extraction equipment, water treatment, etc. In terms of physics, a gas bubble grows with continuously flowing gas while attached along the contact line to the orifice wall before detaching, and freely rises in the liquid after the buoyancy force overcomes the capillary force. The size of detached bubble, however, can be influenced by many factors besides the balance of buoyancy and surface tension such as gas flow rate, orifice size, wetting property of liquid on the orifice wall, and so forth. The bubble formation process often involves complicated fluid dynamics with free surface deformation, disintegration, coalescence, as well as dynamic contact line motion. Its practical importance as well as theoretical modeling challenge has inspired numerous publications by generations of authors (cf. Harkins and Brown 1919; Clift et al. 1978; Gaddis and Vogelpohl 1986; Oguz and Prosperetti 1993; Ponter and Surati 1997; Kulkarni and Joshi 2005; Buwa et al. 2007; Ohta et al. 2011; Simmons et al. 2015). Along with ample experimental results, most theoretical models developed before 1980 were simplified based on force balances with spherically symmetric growing bubbles often assumed as a quasi-static process at very low gas flow rate (as commented by Oguz and Prosperetti 1993; Buwa et al. 2007). To date, even with the available modern experimental equipment and numerical techniques, the general applicability of individually derived theoretical formulas remain elusive (Bari and Robinson, 2013). Due to the complexity of the dynamic process involved in bubble formation, improved understanding of the fundamental physics may only be achieved with more gap-filling model investigations as well as redundant comparisons and cross-examinations.

To simulate the mathematically difficult free-boundary fluid dynamics with minimal simplifying assumptions, various numerical methods for computational fluid dynamice (CFD) have been developed with the advent of powerful computers. For bubbles forming in inviscid or highly viscous liquids governed by linear field equations, a boundary-element (or boundary-integral) method (BEM) in terms of Green's functions with discretization only along the boundaries has been quite effective with results in apparent agreement with some experiments (e.g., Oguz and Prosperetti 1993; Wong et al. 1998; Higuera 2005). The restriction to linear field equations can be eliminated by solving the full nonlinear Navier-Stokes equations with tessellation of entire problem domain, which led to the development of several versions of arbitrary Lagrangian-Eulerian (ALE) methods capable of dealing with large deformations of boundary shapes governed by nonlinear field equations (e.g., Kistler and Scriven 1983; Christodoulou and Scriven 1992) as exemplified in computing highly deformed bubbles (Feng 2007) and bubble formation dynamics up to neck pinch-off (Simmons et al. 2015). However, the difficulties to compute the situation with free surface disintegration and coalescence with the ALE methods have motivated usage of the purely Eulerian methods such as volume of fluid (VOF by Hirt and Nichols 1981), level set method (LSM by Sussman et al. 1998), coupled level set and volume of fluid (CLSVOF by Sussman and Puckett 2000), for investigations of bubble dynamics (e.g, Krishna et al. 1999; Buwa et al. 2007; Nichita et al. 2010; Ohta et al. 2011; Sudeepta Vaishnavi et al. 2023). Unlike ALE with boundary-fitting moving meshes able to track free surface accurately, the resolution of free surface position in VOF simulations depends directly on the Eulerian mesh size; therefore, accurate free surface shape in VOF computation can usually be determined at the expense of a much finer mesh than typical ALE meshes.

Although effective numerical methodologies appear to be well established, most computational codes used in publications are not easily accessible. Code development from scratch even with well-known methodologies can be very resource demanding, and almost prohibitive to most small organization researchers. Fortunately, several open source software packages have been made available in recent years with computational capabilities comparable, or even superior, to those popular commercial packages. Among others, the OpenFOAM® package consists of a variety of common CFD solvers and boundary conditions, with powerful meshing utilities and postprocessing capabilities; it allows user to freely download the C++ source code for close examination and custom modifications, with a large user base across engineering and scientific communities for convenient exchanging knowledge. Hence, a recent version (v2206) of OpenFOAM® VOF solver is selected for self-consistent simulations of the bubble formation process in this work.

Besides providing a gap-filling model investigation, the present monograph with a systematic analysis of tabulated computational data for a set of practically representative cases could also serve as benchmarks for future research comparison. In what follows, section 2 concisely describes the computational model setup, with computational results for a set of case studies presented in section 3. Finally, section 4 provides concluding remarks, highlighting the important findings of the present work.

## 2 Computational Model Description

Considered here is a gas with a constant flow rate through an inlet tube connected to an orifice flowing into a liquid tank, as shown in figure 1a. The two-phase fluid flow problem is assumed axisymmetric, enabling the use of a wedge type mesh (figure 1b) generated with the OpenFOAM® meshing utilities such as blockMesh, mergeMeshes and stitichMesh for combining fine and coarse mesh patches (along the dashed lines) to reduce computational burden while the free surface location can be adequately resolved (e.g., with 40 grid points per orifice diameter around free surfaces, finer than 32 grid points suggested by Ohta et al. 2011, to ensure the variations of computed results within a few percent, as verified by further refining the mesh to 60 grid points per orifice diameter.)

The chosen solver for this type of two-phase flow problem is named interIsoFoam—an improved geometric VOF solver detailed by Roenby et al. (2016), Gamet et al. (2020), and reviewed by Mulbah et al. (2022) as a scheme for high accuracy at low computational cost. Because the description of this type of VOF solver is available in published literature, it is unnecessary to repeat the details here. Briefly in simple terms, the VOF method treats the fluid interface as the "jump" of a scalar phase fraction field from 0 to 1, located in those cells containing fraction values (other than 0 and 1) in an Eulerian mesh. Movements of the scalar phase fraction are governed by a transport equation describing the advection of it with the flow. Then, the traction boundary condition at the free interface is transformed into a body force term associated with the gradient of the scalar phase fraction field in the Navier-Stokes momentum equations in an Eulerian frame, eliminating the need to deal with a moving mesh. It especially makes the simulation of bubble-detaching dynamics much less challenging.

In the present model, the nominal values for liquid phase density and dynamic viscosity are $\rho = 1000$ kg m$^{-3}$ and $\mu = 0.001$ kg m$^{-1}$ s$^{-1}$ (i.e., 1 cp), whereas those corresponding to the gas phase are 1.2 kg m$^{-3}$ and $1.8\times10^{-5}$ kg m$^{-1}$ s$^{-1}$ (0.018 cp), respectively specified for the VOF computations. Over orders-of-magnitude contrasts between liquid and gas phases in terms of these corresponding values would make the influence of gas-phase to the general fluid dynamical behavior almost irrelevant; therefore, only liquid density and viscosity are considered in the analysis and discussion. The value of surface tension at gas-liquid interface is taken as $\sigma = 0.07$ kg s$^{-2}$, closely representing the air-water situation.

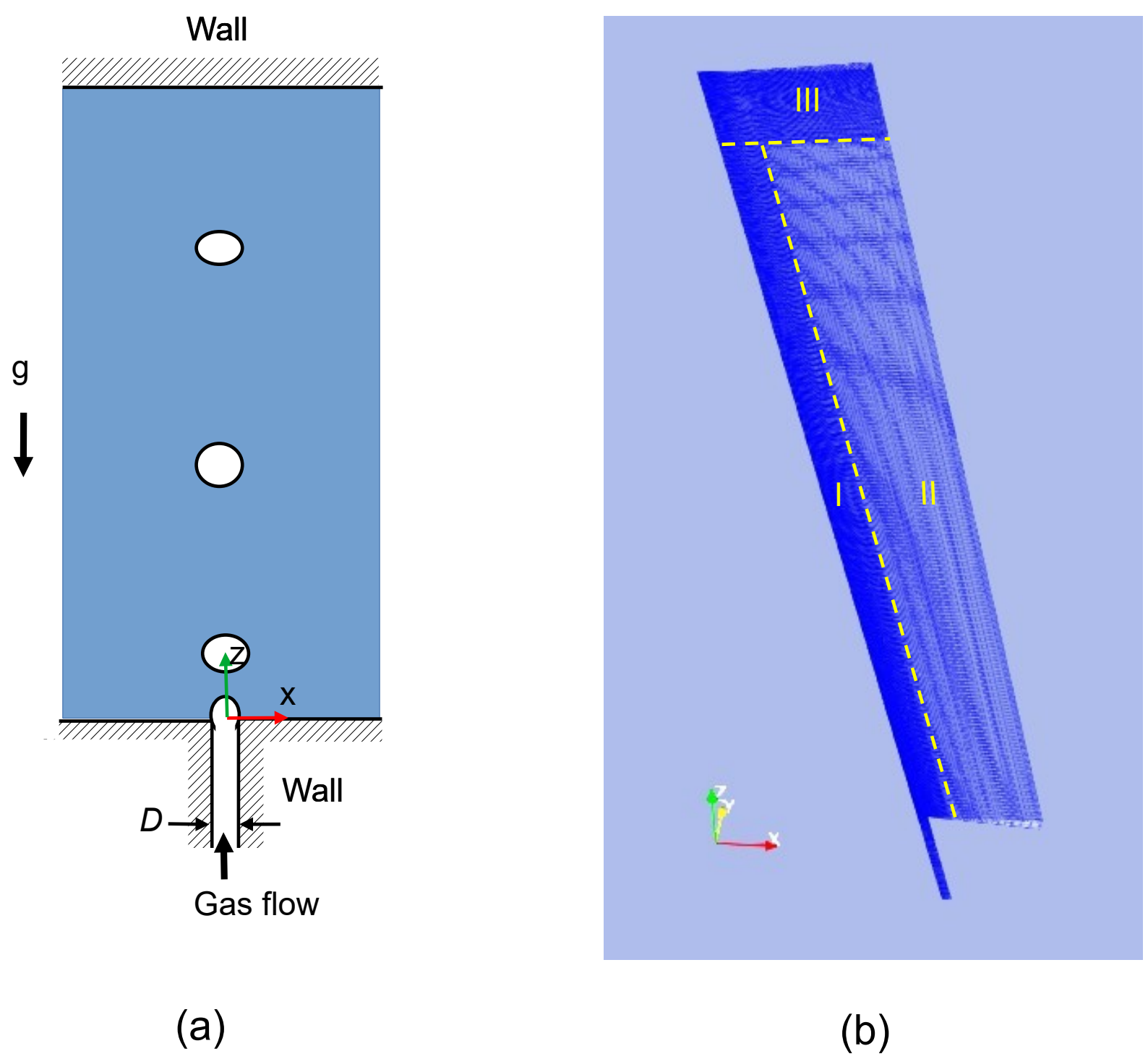

Figure 1. (a) Schematic of bubble formation when a gas flowing through an orifice into a liquid tank; (b) A wedge mesh (discretized into quadrilateral finite-volume cells in two-dimensional space) with a combination of fine (mesh I, mesh III) and coarse (mesh II) patches of about 100,000 total grid points, used for simulations of the axisymmetric two-phase flows in the present bubble formation problem. The top boundary of the computational domain is located at 33 $\times$ $D$ from the orifice, such that its effect on bubble formation becomes inconsequential and is treated as a wall for computational simplicity.

To complete the case specification, the boundary conditions used at the gas inlet are the uniform fixed value type for the velocity vector $U = (0, 0, U_z)$ with $U_z$ = constant and zeroGradient for the piezometric pressure $p$ (cf. description in Feng 2017), at solid walls the noslip type for $U$ and zeroGradient for $p$ , at the open lateral boundary a pressureInletOutletVelocity condition for $U$ and a fixed value type for $p$ (e.g., = $10^5$ Pa). At the gas-liquid-solid three phase contact line, the available dynamicAlphaContactAngle condition is applied as is by specifying the static contact angle $\theta_0$ , advancing contact angle $\theta_A$ and receding contact angle $\theta_R$ as well as a $u_\theta$ value. Here, for a specified static contact angle $\theta_0$ the advancing contact angle and receding contact angle are always assigned as $\theta_A = \theta_0 + 15°$ and $\theta_R = \theta_0 - 15°$ with a fixed value of $u_\theta$ (i.e., = 1 m $s^{-1}$) to keep the presentation logically streamlined and self-consistent. Without losing generality, computed cases in the present work would have a nominal contact angle $\theta_0 = 45^{\circ}$ (corresponding to a partially wetting liquid at the solid wall) unless otherwise specified.

Per OpenFOAM® solver development requirements, each solver must come with at least one tutorial case. In fact, many popular solvers could have multiple tutorial cases provided with the standard downloadable packages. This can be very convenient for users who would not want to spend much time in deciphering the exact meanings of those numerical scheme parameters; some of those tutorial parameter settings can indeed be directly used for similar computational problems without need for modifications. For example, the numerical scheme parameters used in the present work for fvScheme and fvSolution are directly copied from the interIsoFoam tutorial case of "damBreak", using the typical Euler implicit time scheme for temporal integration with adjustable time step under the restriction of Courant number < 0.2.

## 3 Results and Discussion

Before computing numerical solutions, it is useful to estimate a few reference parameters often used in the literature (e.g., Oguz and Prosperetti 1993; Ohta et al. 2011; Simmons et al. 2015). According to a force balance between surface tension and buoyancy (Fritz 1935; van Krevelen and Hoftijzer 1950; Kumar and Kuloor 1970), the Fritz bubble volume $V_F$ and corresponding bubble diameter $d_F$ may be estimated with simple equations,

$$\sigma \pi d_c = \rho g V_F = \frac{\rho g \pi d_F^3}{6} \text{ , thus } V_F = \frac{\sigma \pi d_c}{\rho g} \text{ and } d_F = \left(\frac{6 \sigma d_c}{\rho g}\right)^{1/3} , \quad (1)$$

where $\rho$ denotes the mass density of liquid, $\sigma$ surface tension of the gas-liquid interface, $g$ the acceleration of gravity, and $d_c$ the diameter of three-phase contact line at the orifice wall (which is often assumed to be the same as the orifice diameter $D$.)

The analysis of Oguz and Prosperetti (1993) suggested that the Fritz volume and diameter are expected to be only reasonable for the quasi-static situation when the gas flow rate $Q \ll Q_{crit}$ where the critical gas flow rate may be expressed as

$$Q_{crit} = \pi \left(\frac{16}{3 g^2}\right)^{1/6} \left(\frac{\sigma D}{2 \rho}\right)^{5/6} . \quad (2)$$

At high gas flow rates, i.e., $Q > Q_{crit}$ , the bubbles would detach with a volume proportional to $Q^{6/5}$ .

In what follows, the nominal air-water system is examined first, with cases of variations in liquid viscosity, surface tension in the subsequent subsections.

3.1 Air bubbles in water

For air bubbles in water with $\rho = 1000$ kg m$^3$, $\sigma = 0.07$ kg s$^{-2}$, $g = 9.81$ m s$^{-2}$, and $\mu = 0.001$ kg m$^{-1}$ s$^{-1}$ the value of the Ohnesorge number,

$$Oh = \frac{\mu g^{1/4}}{\rho^{1/4} \sigma^{3/4}} , \qquad (3)$$

becomes $2.313 \times 10^{-3}$, suggesting that the liquid viscosity effect in an air-water system is negligible and an inviscid fluid model for bubble formation can yield reasonable results in agreement with experiments (as shown by Oguz and Prosperetti 1993).

An interesting special situation it that of $d_F = d_c$ which yields

$$d_F = d_c = \sqrt{6} \times \sqrt{\frac{\sigma}{\rho g}} = 2.45 \times L_\sigma = 6.54 \times 10^{-3} \text{ m (or 6.54 mm)} , \qquad (4)$$

where $L_\sigma = \sqrt{\sigma / (\rho g)} = 2.67$ mm is regarded as the capillary scaling length (cf. Longuet-Higgins et al. 1991; Bari and Robinson 2013; Simmons et al. 2015). In other words, the bubble diameter in water is expected to be about the same as the contact line diameter at the orifice edge, when the orifice diameter $D \sim 6.54$ mm. One of the cases presented by Oguz and Prosperetti (1993) with computed data as well as images of transient bubble shapes is that with $D = 4$ mm, which is not too far from that given in (4) and could serve as a reference for comparison.

3.1.1 Case of $D = 4$ mm

For orifice diameter $D = 4$ mm, the reference values of $V_F$ and $d_F$ are 89.67 μL (or mm$^3$) and 5.55 mm with $Q_{crit} = 3768.77$ μL/s (assuming $d_c = D$) Figure 2 shows the computed results visualized with ParaView for bubble formation dynamics at various gas flow rates. As indicated by the spacing between the sequential bubbles, the bubble volume is increasing with the gas flow rate because the spacing does not decrease inversely proportional to the gas flow rate. For low gas flow rates $Q < Q_{crit}$ , the detached bubbles rise following one another in a so-called period-1 regime with the bubble volume gradually increasing with $Q$ (as shown in table 1). When $Q \sim Q_{crit}$ (i.e., at $U = 300$ mm/s), the spacing between bubbles become so close that coalescence of two neighboring bubbles occurs and resulting in paring bubbles in the period-2 regime (cf. Buwa et al. 2007). If the gas flow rate were further increased, it would eventually lead to chaos via period doubling (cf. Tufaile and Sartorelli 2000).

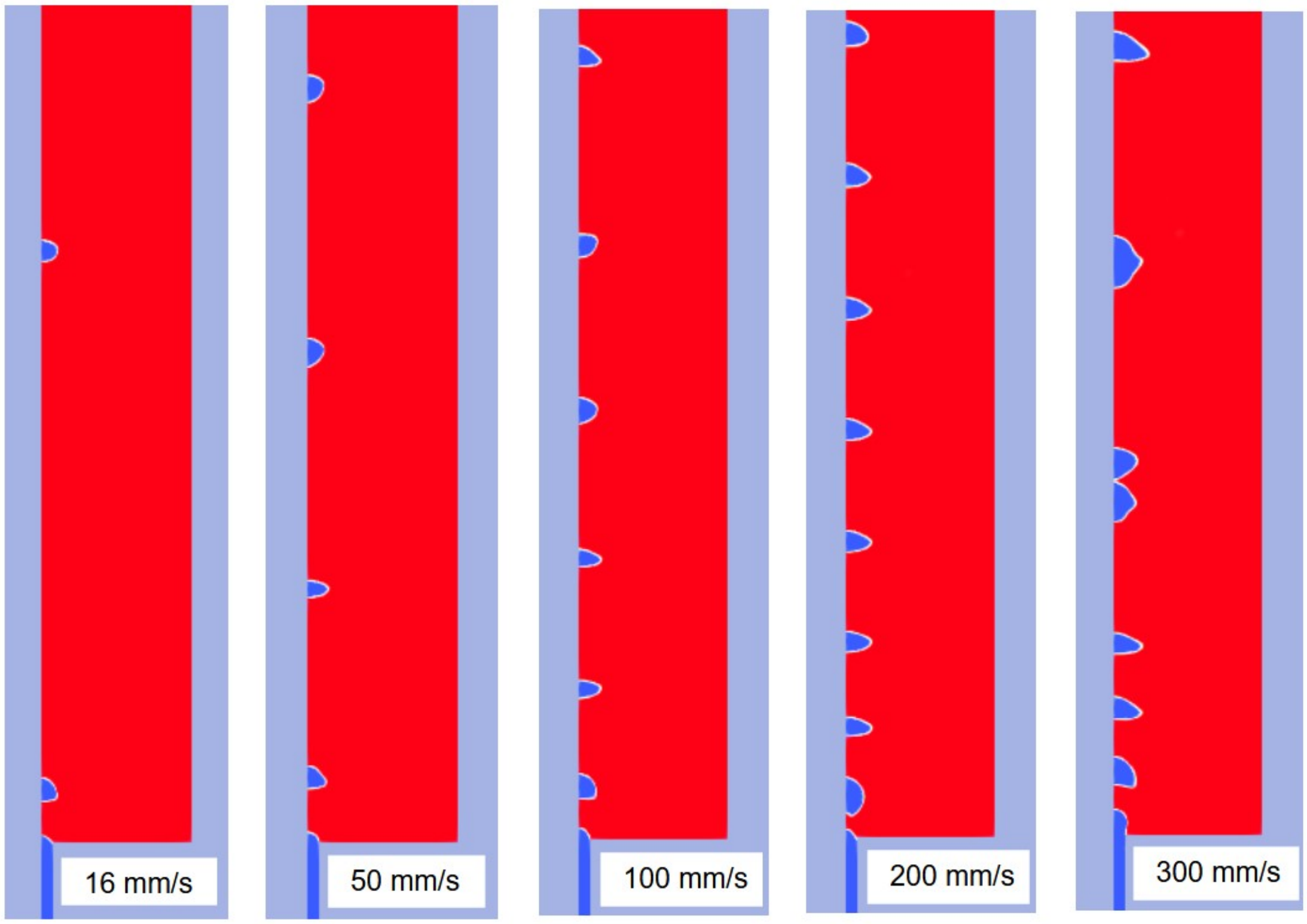

Figure 2. Formation of air bubbles in water from an orifice of diameter $D = 4$ mm at various air flow velocity (i.e. $U = 0.016$, 0.05, 0.1, 0.2, 0.3 m/s corresponding to flow rate $Q = 201$, 628, 1257, 2513, 3770 μL/s), simulated with the OpenFOAM® VOF solver interIsoFoam and visualized with the postprocessor ParaView.

To provide more quantitative information, the computed data of bubble detachment period $\Delta T$ with the corresponding bubble volume $V$ (derived from $V = \Delta T \times Q$ which matches the numerical integration result along the free surface of the detached bubbles) and diameter $d$ are given in table 1, wth a reference to the $Q / Q_{crit}$ value for $D = 4$ mm. In comparison with the results by Oguz and Prosperetti (1993) for $Q = 0.008$ and 0.016 μL/s as $d = 5.188$ and 5.277 mm, the present results of $d = 4.71$ and 4.81 mm are consistently smaller (by about 10%). Despite the numerical differences between those values and that of Oguz and Prosperetti (1993), we are in agreement on the general trend of bubble size decreasing with reducing $Q$. Both of our values of $V$ and $d$ are less than the corresponding $V_F$ (= 89.67 μL) and $d_F$ (= 5.55 mm) from (1) which are usually regarded as the minimum values of bubble volume and diameter at the limit of so-called quasi-static bubble formation (at $Q << Q_{crit}$). The fact that $V_F$ and $d_F$ can be inaccurate in quasi-static treatment were explained theoretically with the dynamic process of necking and pinch-off by Simmons et al. (2015) in a model with pinned contact line. That the Fritz volume $V_F$ needs correction was also recognized in an early publication of Harkins and Brown (1919).

More experimental evidence for $V < V_F$ mostly with $Q / Q_{crit} < 0.1$ has also been shown in publications of Corchero et al. (2006), Bari and Robinson (2013), and Manoharan et al. (2021). According to the fitted power law of Bari and Robinson (2013) for measured quasi-static bubble volume $V_B$ in experiments with fixed contact line:

$$\frac{V_B}{V_F}=0.6863\left(\frac{D}{L_\sigma}\right)^{-0.116} , \tag{5}$$

the value of $V_B$ would become ~54.49 mm$^3$ for $D$ = 4 mm (with the capillary length in (4) $L_\sigma$ = 2.67 mm), in agreement with the value of $V$ in table 1 for $Q << Q_{crit}$ .

Table 1. Computational results of bubble detachment period $\Delta T$, corresponding bubble volume $V$ and bubble diameter $d$ with a reference to the valiue of $Q / Q_{crit}$ for $D$ = 4 mm (at $\theta_0$ = 45°)

| $U$ (m/s) | $Q$ (μL/s) | $\Delta T$ (ms) | $V$ (mm$^3$) | $d$ (mm) | $Q / Q_{crit}$ |
|---|---|---|---|---|---|
| 0.008 | 100.5 | 545 | 54.79 | 4.71 | 0.0266 |
| 0.016 | 201.1 | 289 | 58.11 | 4.81 | 0.0533 |
| 0.05 | 628.3 | 113 | 71.00 | 5.14 | 0.167 |
| 0.1 | 1256.6 | 68 | 85.45 | 5.46 | 0.333 |
| 0.2 | 2513.3 | 50 | 125.66 | 6.21 | 0.667 |
| 0.3 | 3769.9 | 41 | 154,57 | 6.66 | 1.000 |

If the data in table 1 is plotted in terms of $V$ versus $x = Q / Q_{crit}$ (< 1), a fitted line can be obtained as $V(x)$ = 103.84 $x$ + 52.71 mm$^3$ with $R^2 > 0.997$. Thus, the present results suggest a monotonic increase of $V$ with $Q$ and $V(0)$ = 52.71 mm$^3$, unlike what has been claimed by most other authors as the static regime wherein $V$ becomes nearly constant for $Q / Q_{crit} << 1$. According to the data in table 1, the bubble size may become close to those of the Fritz value (1) in the $Q / Q_{crit}$ interval between 0.1 and 0.5. The deviation of $V$ from $V(0)$ would be < 10% for $Q / Q_{crit} < 0.05$.

The transient bubble free surface profiles are shown in figure 3 during the bubble detachment process at $U$ = 0.008, 0.016, 0.1 m/s. A series of bubble surface profiles for $U$ ~ 0.016 m/s ($Q$ ~ 200 μL/s) were presented by Oguz and Prosperetti (1993), qualitatively comparable to the corresponding profiles in figure 3. However, obvious discrepancy exists between their free surface shapes in figure 4 during attached bubble growth at particular time points, especially for $t$ < -10 ms. A closer examination reveals an inconsistency in their reported numerical free surface shapes which were apparently intended for compariing with the experimental images by Longuet-Higgins et al. (1991) for about one bubble per second, likely with $Q < 60$ μL/s rather than $\Delta T \sim 300$ ms with $Q \sim 200$ μL/s. Figure 4 shows that a well-developed neck appears at free surface typically when $t > -10$ ms, thereafter the time that takes for neck pinch-off would be almost independent of air flow rate but controlled by the capillary timescale (cf. Oguz and Prosperetti 1993),

$$t_\sigma = \sqrt{\frac{\rho D^3}{8\sigma}} \ , \qquad (6)$$

which becomes 10.69 ms for $D = 4$ mm (with $\rho = 1000$ kg m$^3$ and $\sigma = 0.07$ kg s$^{-2}$).

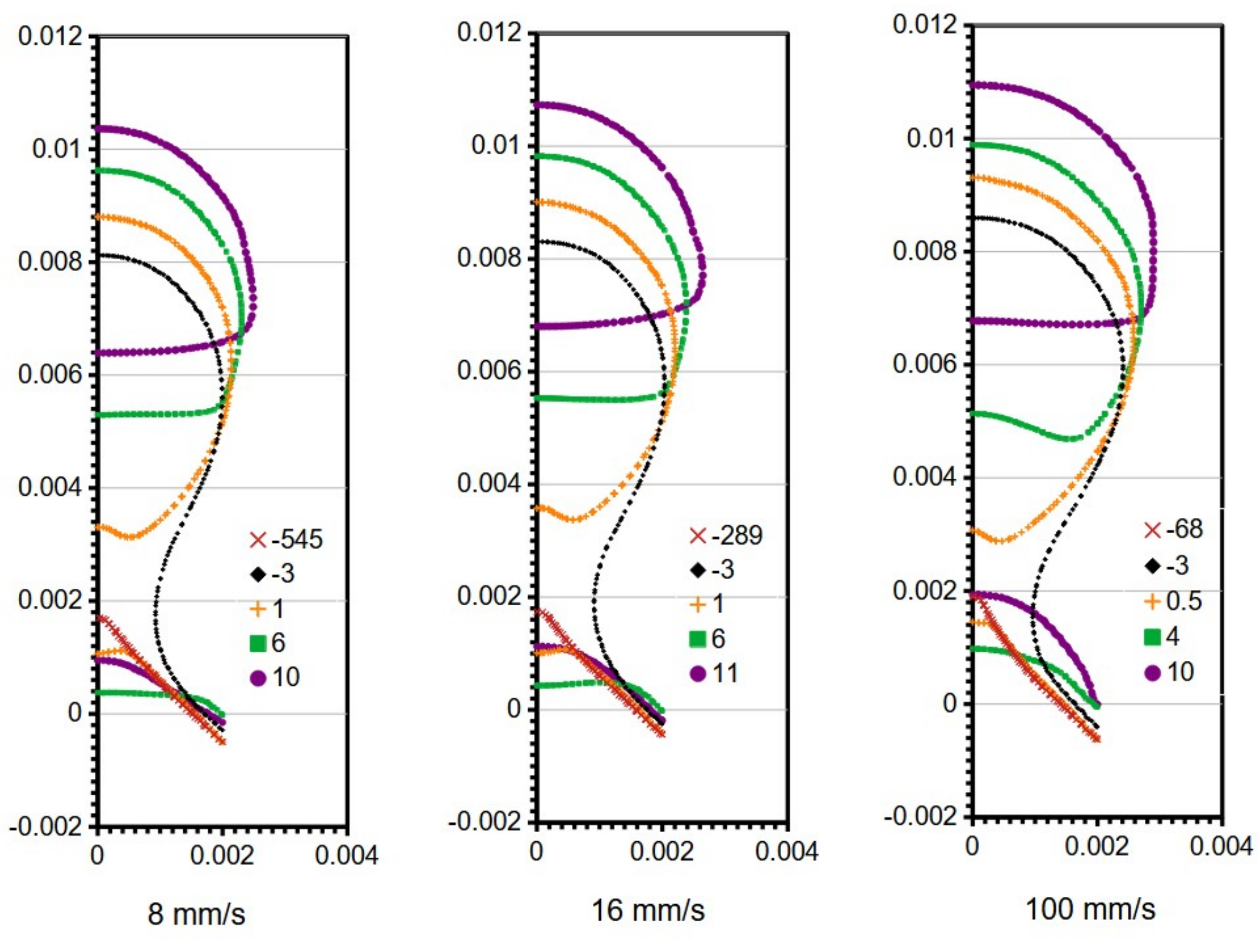

Figure 3. Free surface shapes during air bubble formation from an orifice of diameter $D = 4$ mm in water at $U = 0.008$, 0.016, 0.1 m/s corresponding to $Q = 100.5$, 201, 1257 μL/s.

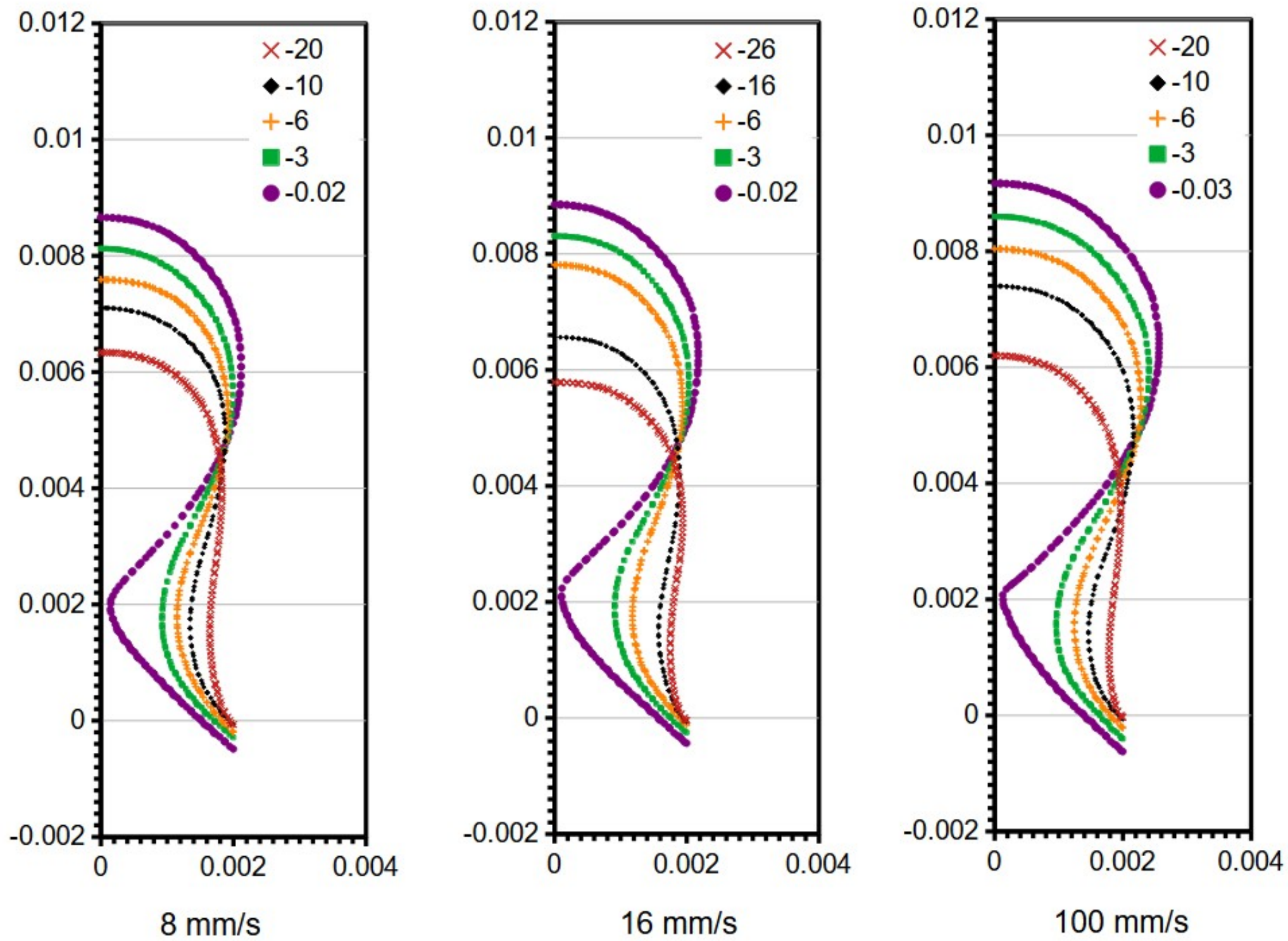

Figure 4. As figure 3 but only for during bubble growth toward necking pinch-off.

Considering the physical mechanisms, the bubble formation process may be divided into two stages (cf. Clift et al. 1978; Simmons et al. 2015): the initial growth governed by the hydrostatic and capillary pressures, followed by a capillary-force driven dynamic event with considerable local velocity during necking and pinch-off. The timescale for the initial growth is expected according to the gas flow rate, namely proportional to $V_F / Q$, while that governing the necking dynamics is likely related to the capillary timescale $t_\sigma$ in (6). At a low gas flow rate with the quasi-static bubble formation process, necking starts before the bubble volume reaches the static force-balance value $V_F$ at a time $\phi_b V_F /Q < V_F / Q$ until the bubble detachment, with a timescale $\phi_a t_\sigma$ independent of $Q$, as evidenced in figure 4. When $\Delta T = (\phi_a t_\sigma) + (\phi_b V_F/Q) < (V_F/Q)$, the detached bubble volume,

$$V = (\phi_a t_\sigma) Q + (\phi_b V_F), \tag{7}$$

should be less than the Fritz volume $V_F$ as observed in Table 1. If both $\phi_a$ and $\phi_b$ are independent of $Q$ , (7) suggests a linear relationship between $V$ and $Q$ instead of a constant, consistent with the physical expectation.

With the default setting of contact angle $\theta_0 = 45^o$ (with $\theta_A = 60^o$ and $\theta_R = 30^o$), figures 3 and 4 also reveal the fact that the three-phase contact line could move along the vertical inner wall of the inlet tube, which apparently has not been recognized in the published literature. The dynamic contact line can reach the deepest location ($r$ = 2 mm, $z$ = -0.43 mm for $U$ = 0.016 m/s) into the inner tube wall as the bubble detachment occurs, while outside the time interval of $-t_\sigma < t < t_\sigma$ (where $t_\sigma \sim 11$ ms) the contact line appears to stay pinned at the orifice edge ($r$ = 2 mm, $z$ = 0). Such a phenomenon of contact line movement seems to be driven by the capillary effects within the capillary timescale.

In view of the fact that Oguz and Prosperetti (1993) used a '90°-rule' for contact angle, a case of $\theta_0 = 90^o$ is computed for $U$ = 0.016 m/s ($Q$ = 201 μL/s). The results show $\Delta T$ = 292 ms with $V$ = 58.71 mm$^3$ and $d$ = 4.82 mm, which are practically no difference from the corresponding values in table 1. However, with $\theta_0 = 90^o$ the contact line would be pinned at the orifice edge all the time (referred to as 'mode A' bubble formation in Gerlach et al. 2005), in contrast to the case of $\theta_0 = 45^o$ in figures 3 and 4. The bubble surface profiles shown by Oguz and Prosperetti (1993) for this case indeed seem to have pinned contact line at the thin-wall needle edge. According to the present results, the detached bubble volume at a given gas flow rate does not seem to vary much when the liquid is adequately wetting the orifice wall with the contact angle less than that for contact line pinning at the orifice edge, because the contact diameter $d_c$ remains the same as $D$ regardless how the contact line moves on the inlet tube wall.

However, if the contact angle is increased to $\theta_0 = 120^o$ with $U$ = 0.016 m/s ($Q / Q_{crit}$ = 0.0533), we would now obtain $\Delta T$ = 483 ms with $V$ = 97.11 mm$^3$ and $d$ = 5.70 mm, while the contact line could reach as far as $r \sim 4$ mm along the horizontal surface of $z$ = 0 during the time outside the interval ($-t_\sigma$ , $t_\sigma$). For $\theta_0 = 105^o$, we then obtain $\Delta T$ = 347 ms with $V$ = 69.77 mm$^3$ and $d$ = 5.11 mm, also with contact line moving on the horizontal outside wall of orifice. Actually, the contact line could move on the horizontal surface when $\theta_0 > 95^o$ (for $\theta_0 = 95^o$, the contact line appears to be barely pinned at orifice edge and the value of $\Delta T$ is within a few percent of that for $\theta_0 = 45^o, 90^o$). The situation when the contact line moves on the horizontal outside wall of orifice during bubble formation is referred to as 'mode B' in Gerlach et al. (2005). In mode B the bubble volume was shown to increase with the contact angle almost exponentially from $\theta_0 \sim 50^o$ to 110° (Gerlach et al. 2007), because the time-dependent contact diameter $d_c$ is always greater than the orifice diameter $D$.

For a given geometric setting, there exists a critical contact angle $\theta_c$ (e.g., ~ 95° for $D$ = 4 mm at $U$ = 0.016 m/s) dividing the Mode A and Mode B regimes, namely below $\theta_c$ the contact line could move along the vertical inner wall of inlet channel whereas above $\theta_c$ the contact line is moving on the horizontal outside wall of orifice. The motion of contact line along the vertical inner wall has little effect on detached bubble size; but the bubble size can be very sensitive to the contact line motion on the horizontal outside wall of orifice due to transient variation of the contact diameter. The fact that below a critical contact angle $\theta_c$ the detached bubble size appears independent of the contact angle was also reported by Manoharan et al. (2021) with their experiments.

3.1.2 Case of $D$ = 2 mm

For $D$ = 2 mm, the reference values of $V_F$ and $d_F$ are 44.83 μL and 4.41 mm with $Q_{crit}$ = 2115.15 μL/s and $t_\sigma$ = 3.78 ms. To compare with a result of $d$ = 4.40 mm for $Q$ ~ 391 μL/s by Oguz and Prosperetti (1993), a case of $\theta_0$ = 90º is computed for $U$ = 0.125 m/s ($Q$ = 392.70 μL/s) which yields $\Delta T$ = 147 ms with $V$ = 57.73 mm$^3$ and $d$ = 4.80 mm. For $\theta_0$ = 80º, $\Delta T$ = 101 ms with $V$ = 39.66 mm$^3$ and $d$ = 4.23 mm. However, with the present model, the contact line would move on the horizontal surface for $\theta_0$ > 60º (at $\theta_0 = \theta_c$ ~ 60º the contact line appears barely pinned at the orifice edge with computed $\Delta T$ = 80 ms). In view of $\theta_c$ ~ 95º for $D$ = 4 mm, the computed case of $D$ = 2 mm suggests a decreasing trend of $\theta_{crit}$ with reducing orifice size. As a reference, the value of $V_B$ from (5) (for $\theta_0 < \theta_c$) would become ~29.53 mm$^3$ for $D$ = 2 mm at low gas flow rates for quasi-static bubble growth.

Table 2 presents the computed results for $Q$ ~ 393 and 1668 μL/s each at $\theta_0$ = 45º and 75º, respectively. The former is intended to compare with a case reported by Oguz and Prosperetti (1993). while the latter with some results of Gerlach et al. (2007) who reported $\Delta T$ = 42, 54, 126 ms for $\theta_0$ = 0º, 70º, 110º, respectively at $Q$ = 1667 μL/s (100 mL/min) with their numerical simulations using a static contact angle model (i.e., for $\theta_A = \theta_R = \theta_0$).

Table 2. As table 1 but for $D$ = 2 mm with $V_F$ = 44.83 μL, $d_F$ = 4.41 mm, $Q_{crit}$ = 2115.15 μL/s.

| $U$ (m/s) | $Q$ (μL/s) | $\theta_0$ (deg) | $\Delta T$ (ms) | $V$ (mm$^3$) | $d$ (mm) | $Q$ / $Q_{crit}$ |
|---|---|---|---|---|---|---|
| 0.125 | 392.70 | 45 | 80 | 31.42 | 3.91 | 0.186 |
| | | 75 | 93 | 36.52 | 4.12 | |
| 0.531 | 1668.18 | 45 | 36 | 60.05 | 4.86 | 0.789 |
| | | 75 | 47 | 78.40 | 5.31 | |

Apparently, the present results consistently show smaller bubble volume than those reported by both Oguz and Prosperetti (1993) and Gerlach et al. (2007). Although the governing equations for the bubble formation problem are mostly the same, the treatments of dynamic behavior of the contact line are noticeably different between the present model and those previous authors. Before resolving the exact causes (which can be extremely difficult due to inaccessible source codes), the differences beyond normally acceptable numerical tolerance may be attributed to the differences in mathematical representation of dynamic contact line, for now.

Nonetheless, the present results are in qualitative agreement with the previous publications with the same trends for variations of wetting angle and gas flow rate.

To provide a reference for the "quasi-static regime" at low gas flow rate (where the bubble size is expected to vary little), a case of $U = 0.05$ m/s ($Q / Q_{crit} = 0.0743$) is computed yielding bubble diameters $d = 3.81$ mm corresponding to a volume $V = 28.90$ mm$^3$ for $\Delta T = 184$ ms with $\theta_0 = 45^o$ (which is practically the same as that up to $\theta_0 = 75^o$). Now, the critical contact angle becomes $\theta_c \sim 75^o$ (for the contact line pinned at the orifice edge), indicating an increase of $\theta_c$ with reducing the gas flow rate. But with $\theta_0 = 90^o$ significant contact line motion on horizontal surface occurs, as a consequence the computed $d = 4.59$ mm, $V = 50.74$ mm$^3$ with $\Delta T = 323$ ms. Actually the contact line motion on horizontal outside wall of orifice becomes obvious at $\theta_0 = 80^o$ with $\Delta T = 219$ ms, $V = 34.40$ mm$^3$ and $d = 4.04$ mm. Again, the present results for partially wetting liquids with $\theta_0 < \theta_c$ are in good agreement with the fixed contact line value of $V_B \sim 29.53$ mm$^3$ from (5) for $D = 2$ mm.

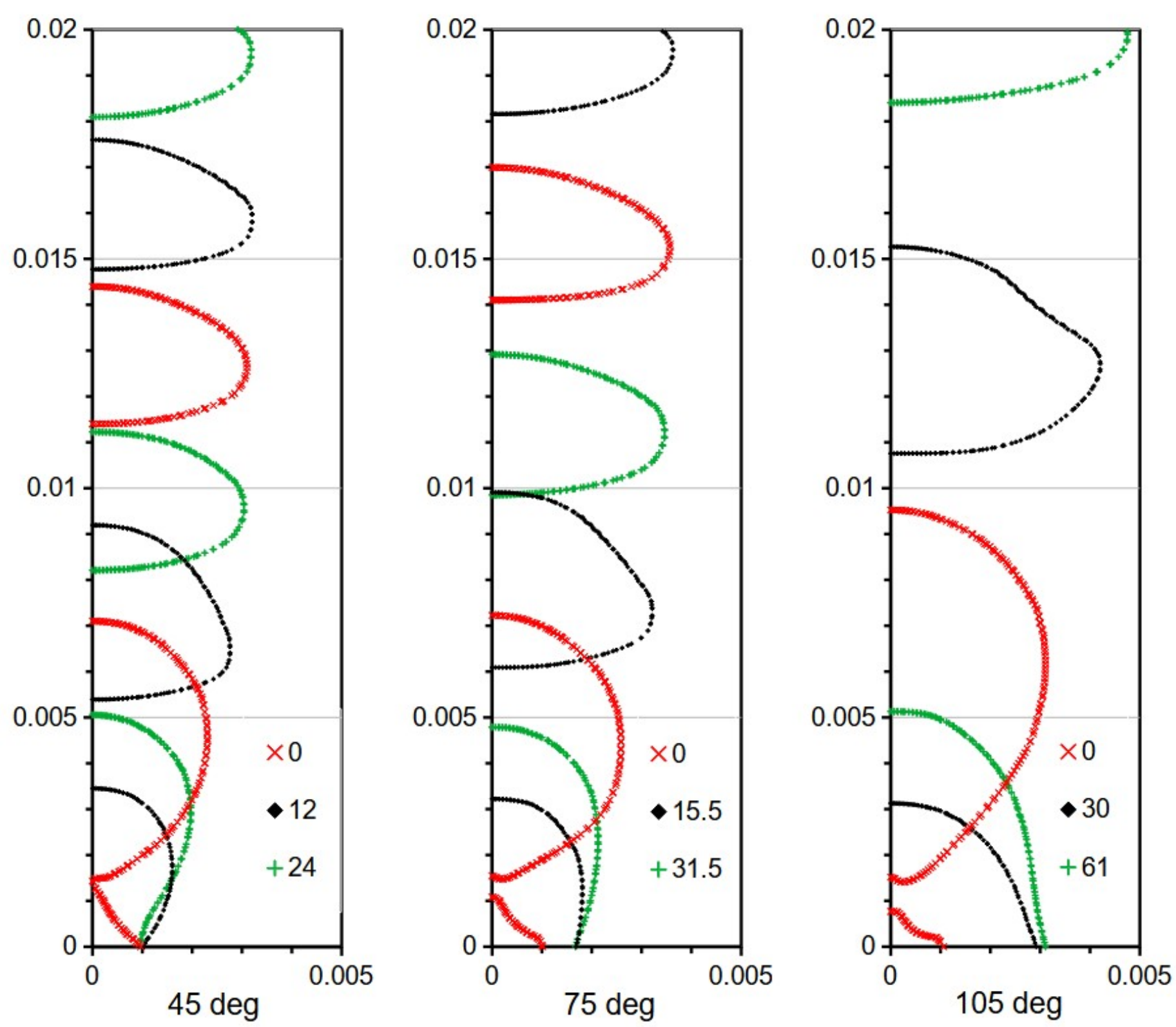

Figure 5. As figure 3 but for $D = 2$ mm at $U = 0.531$ m/s corresponding to flow rate $Q = 1668$ μL/s for $\theta_0 = 45^o, 75^o, 105^o$ (with $\theta_A = \theta_0 + 15^o$ and $\theta_R = \theta_0 - 15^o$) and $\Delta T = 36, 47, 91$ ms.

Further comparison with some computed bubble surface profiles in Gerlach et al. (2007) for $\theta_0$ = 0º, 70º, and 110º at $Q$ = 1667 µL/s and $t$ = 0, $\Delta T/3$, 2 $\Delta T/3$ is provided in figure 5 as present computational results for $\theta_0$ = 45º, 75º, and 105º exhibiting similar contact line motion on the horizontal surface (which could occur for $\theta_0$ > 60º per present computations). The bubble surface profiles in figure 5 look generally comparable to those corresponding ones of Gerlach et al. (2007).

For the record, a fitted line can be obtained as $V(x) = 44.24\, x + 24.49$ mm$^3$ with $R^2 >$ 0.997 for $\theta_0$ = 45º and $D$ = 2 mm with computed data for $x = Q / Q_{crit}$ from 0.037 up to 1.04. This is consistent with (7) having $\phi_a\, t_\sigma\, Q_{crit}$ = 44.24 mm$^3$ and $\phi_b\, V_F$ = 24.49 mm$^3$ corresponding to $\phi_a \sim$ 5.53 and $\phi_b \sim$ 0.55. Unlike that seen in figure 2 for $D$ = 4 mm with pairing bubbles at $Q / Q_{crit}$ = 1, here for $D$ = 2 mm the detached bubbles remain period-1 when $Q / Q_{crit}$ = 1.04 (with pairing bubbles observed at $Q / Q_{crit}$ = 1.485 as slight movements of the contact line can be noticed on the horizontal outside wall of orifice, consistent with the decreasing trend of $\theta_c$ with the gas flow rate $Q$).

### 3.1.3 Case of $D$ = 1.5 mm

In a study of the wetting effects on bubble formation, Manoharan et al. (2021) presented experimental results for a list of averaged bubble diameters detached from $D$ = 1.5 mm with different contact angles with respect to water, in the quasi-static regime (for $Q$ < 150 µL/s or < 9 mL/min such that $Q / Q_{crit}$ < 0.09): $d$ = 2.75, 2.69, 2.98, 3.78, 3.98, 4.27 mm with $\theta_0$ = 43º, 52º, 62º, 105º, 120º, 128º (in their Table 2). For $D$ = 1.5 mm, we would have $V_F$ = 33.63 µL and $d_F$ = 4.00 mm with $Q_{crit}$ = 1664.28 µL/s and $t_\sigma$ = 2.45 ms. Yet the value of $d$ = 3.52 mm corresponding to $V_B$ = 22.90 mm$^3$ from (5), which seems in reasonable agreement with the present computational results in table 3 for $\theta_0 < \theta_c \sim 65º$ . Noteworthy here is that at $\theta_0$ = 70º even with noticeable contact line motion on the horizontal outside wall of orifice the computed bubble size does not seem to change from that for $\theta_0$ < 65º (with $Q / Q_{crit}$ = 0.053 corresponding to $U$ = 0.05 m/s). In this case, the contact line motion on the horizontal outside wall of orifice occurs only when ~15 ms away from the time of meniscus pinch-off, much longer than $t_\sigma$ = 2.45 ms.

It is interesting to note that the bubble sizes in table 3 are consistently greater than those corresponding ones shown by Manoharan et al. (2021), unlike other comparisons with previously published data. However, the present results for $\theta_0$ < 75º are in quite good agreement with the value of $V_B$ according to (5) based on the experiments of Bari and Robinson (2013) for quasi-static bubble growth with fixed contact line.

Table 3. As table 1 but for $D$ = 1.5 mm with $V_F$ = 33.63 μL, $d_F$ = 4.00 mm, $Q_{crit}$ = 1664.28 μL/s.

| $U$ (m/s) | $Q$ (μL/s) | $\theta_0$ (deg) | $\Delta T$ (ms) | $V$ (mm$^3$) | $d$ (mm) | $Q$ / $Q_{crit}$ |
|---|---|---|---|---|---|---|
| 0.05 | 88.36 | 45 | 218 | 19.26 | 3.32 | 0.053 |
| | | 60 | 214 | 18.91 | 3.31 | |
| | | 75 | 283 | 25.01 | 3.63 | |
| | | 90 | 540 | 47.71 | 4.50 | |
| | | 105 | 888 | 78.46 | 5.31 | |
| | | 120 | 884 | 78.11 | 5.30 | |

3.1.4 Case of $D$ = 1 mm

Reducing orifice size to $D$ = 1 mm would have $V_F$ = 22.42 μL and $d_F$ = 3.50 mm with $Q_{crit}$ = 1187.09 μL/s and $t_\sigma$ = 1.34 ms, while $V_B$ = 16.00 μL (corresponding to $d$ = 3.13 mm) from (5). Table 4 shows the computed results for $Q$ / $Q_{crit}$ = 0.033 and at various $\theta_0$ with $\theta_c < 60^o$ (at $\theta_0 = 60^o$ the contact line motion on horizontal outside wall of orifice is noticeable although the detached bubble volume does not seems to increase). Noteworthy here is that at $\theta_0 = 65^o$ with substantial contact line movements on horizontal outside wall of orifice, the detached bubble volume remains comparable to that at $\theta_0 < \theta_c$ . Even up to $\theta_0 = 70^o$ the detached bubble volume still remains about the same as $V_B$ (though larger than that with $\theta_0 = 65^o$ for quasi-static bubble formation based on the present computational model).

Some experimental results of bubble formation in water for $D$ = 1 mm at $Q$ = 48.4 μL/s ($U$ ~ 0.0616 m/s) with $\theta_0$ = 68$^o$ were presented by Corchero et al. (2006) yielding bubbles of $V$ ~ 22.4 mm$^3$ and $d$ ~ 3.5 mm. This is quite comparable to $V_F$ and $d_F$ which would happen at $\theta_0$ ~ 75$^o$ according to present results shown in table 4. Imperfect quantitative comparison between the computational model and experiment may not be surprising in view of the intrinsic limitations of macroscopic continuum models for the dynamic contact line (cf. Blake and Ruschak 1997; Chen et al. 2009) as well as difficulties in experimentally measuring the moving contact line behavior. Based on the present computational findings, it is clear that the preferred comparison cases would be those with $\theta_0 < \theta_c$ where the detached bubble sizes are insensitive to the actual $\theta_0$ values. The situation of $\theta_0 < \theta_c$ could also be practically desirable for minimizing the bubble size with improved process consistency. However, model prediction of the value of $\theta_c$ for contact line pinning is likely to depend on dynamic contact line formulation and the actual value of $\theta_c$ in experiments can be sensitive to the surface properties of the orifice materials.

Table 4. As table 1 but for $D$ = 1 mm with $V_F$ = 22.42 μL, $d_F$ = 3.50 mm, $Q_{crit}$ = 1187.09 μL/s.

| $U$ (m/s) | $Q$ (μL/s) | $\theta_0$ (deg) | $\Delta T$ (ms) | $V$ (mm$^3$) | $d$ (mm) | $Q$ / $Q_{crit}$ |
|---|---|---|---|---|---|---|
| 0.05 | 39.27 | 45 | 274 | 10.76 | 2.74 | 0.033 |
| | | 65 | 282 | 11.07 | 2.77 | |
| | | 70 | 419 | 16.45 | 3.16 | |
| | | 75 | 586 | 23.01 | 3.53 | |

To examine the effect of gas flow rate $Q$ for $\theta_0 < \theta_c$ cases at $\theta_0$ = 45° with $Q$ = 392.7 μL/s ($U$ = 0.5 m/s, $Q$ / $Q_{crit}$ = 0.331), $Q$ = 1178.1 μL/s ($U$ = 1.5 m/s, $Q$ / $Q_{crit}$ = 0.992), and $Q$ = 1570.8 μL/s ($U$ = 2 m/s, $Q$ / $Q_{crit}$ = 1.323) are also computed here. As consistent with that already being observed, increasing $Q$ / $Q_{crit}$ to > 0.992 would reduce the value of $\theta_c$ to < 45° , because now the contact line motion at $\theta_0$ = 45° is noticeable on the horizontal outside wall of orifice.

Again, the computed data at $Q$ / $Q_{crit}$ = 0.033, 0.331, 0.992, 1.323 for $\theta_0$ = 45° can be fitted into a line of $V$ versus $x$ = $Q$ / $Q_{crit}$ as $V(x)$ = 21.45 $x$ + 9.179 mm$^3$ with $R^2$ > 0.995 in the form of (7) having $\phi_a$ ~ 13.52 and $\phi_b$ ~ 0.41. Comparing the values of $\phi_a$ and $\phi_b$ for the case of $D$ = 2 mm, the differences suggest that $\phi_a$ and $\phi_b$ are at least orifice-size dependent. The phenomenon of bubble pairing for an orifice of $D$ = 1 mm with $\theta_0$ = 45° does not seem to occur until $Q$ / $Q_{crit}$ > 2, and definitely occurs when $U$ = 3.5 m/s corresponding to $Q$ / $Q_{crit}$ = 2.32 with $\Delta T$ = 23.6 ms, $V$ = 64.87 mm$^3$ and $d$ = 4.99 mm (which is above but not too far from the value the fitted linear formula $V$(2.32) = 58.85 mm$^3$). It should be noted that the value of $V$ here is obtained from neck pinch-off period $\Delta T$ for individual bubbles before pairing coalescence. Apparently, the value of $Q$ / $Q_{crit}$ for pairing bubbles increases with reducing the orifice size.

Moreover, Mirsandi et al. (2020) showed experimental as well as computational results for $D$ = 1 mm around $Q$ = 21.8 μL/min (i.e., 363 μL/s and $U$ ~ 0.462 m/s) with $\theta_0$ ~ 105° and $\Delta T$ ~ 351 ms, apparently corresponding to bubbles of $V$ ~ 127 mm$^3$ and $d$ ~ 6.2 mm. For a comparison, a case of $U$ = 0.462 m/s at $\theta_0$ ~ 105° (with $\theta_A$ = 120° and $\theta_R$ = 90°) is computed here yielding bubbles of $V$ = 87.08 mm$^3$ and $d$ = 5.50 mm, significantly larger than that with $\theta_0$ = 45° (i.e., having $V$(0.306) ~ 15.5 mm$^3$) but not quite as large as that of Mirsandi et al. If the value of $D$ were increased to 2 mm and 4 mm with all other parameters kept the same (for $Q$ ~ 363 μL/s), the resulting $d$ would respectively become 5.63 and 5.57 mm, not really much different from $d$ = 5.50 mm with $D$ = 1 mm. Then, for non-wetting liquids with respect to the orifice wall, the bubble size may not decrease with reducing the orifice size at a given gas flow rate.

3.2 Air bubbles in an aqueous-glycerol solution

For air bubbles forming from an orifice of $D$ = 1.7 mm in an aqueous-glycerol solution with $\rho$ = 1223.8 kg m$^{-3}$, $\sigma$ = 0.066 kg s$^{-2}$, and $\mu$ = 0.126 kg m$^{-1}$ s$^{-1}$ (Helsby and Tuson 1955; Ohta et al. 2011), the values of $V_F$ and $d_F$ become 29.36 µL and 3.83 mm. According to (3) the Ohnesorge number in this case would take a value of $Oh$ = 0.2895, much larger than 2.313×10$^{-3}$ for the air-water system. This is an exemplifying case for the effect of increasing viscosity by two orders of magnitude.

Table 5. As table 1 but for air bubbles in an aqueous-glycerol solution with $D$ = 1.7 mm, $V_F$ = 29.36 µL, $d_F$ = 3.83 mm, $Q_{crit}$ = 1486.40 µL/s

| $U$ (m/s) | $Q$ (µL/s) | $\theta_0$ (deg) | $\Delta T$ (ms) | $V$ (mm$^3$) | $d$ (mm) | $Q$ / $Q_{crit}$ |
|---|---|---|---|---|---|---|
| 0.088 | 199.74 | 45 | 140.5 | 28.06 | 3.77 | 0.134 |
| | | 75 | 171 | 34.16 | 4.03 | |
| 0.441 | 1000.98 | 45 | 60.5 | 60.56 | 4.87 | 0.673 |
| | | 75 | 72 | 72.07 | 5.16 | |
| 0.881 | 1999.69 | 45 | 48.2 | 96.39 | 5.69 | 1.345 |
| | | 75 | 56.0 | 111,98 | 5.98 | |

Table 5 shows the present model results for $U$ = 0.088, 0.441, 0.881 m/s at $\theta_0$ = 45° and 75°, for comparing with a set of computational results reported by Ohta et al. (2011), i.e., $d$ = 4.22, 4.89, 5.49 mm, and corresponding experimental results of Helsby and Tuson (1955) as $d$ = 4.06, 499, 572 mm. In view of the fact that both Helsby and Tuson (1955) and Ohta et al. (2011) were considering bubble detaching from a thin-wall needle rather than an orifice on a plane surface, their results should be reasonably described with the present model for cases of $\theta_0$ = 45° with contact diameter remaining constant. But the cases of $\theta_0$ = 75° are also computed here for references. Indeed, the present computations for $\theta_0$ = 45° yield results in good agreement with Helsby and Tuson (1955) as well as Ohta et al. (2011), although for the case of $U$ = 0.088 m/s the case of $\theta_0$ = 75° seems to offer even better agreement. Interestingly, the present values of $V$ = 28.06 mm$^3$ and $d$ = 3.77 mm at $Q$ / $Q_{crit}$ = 0.134 are rather close to $V_F$ = 29.36 µL and $d_F$ = 3.83 mm than those of Helsby & Tuson (1955) and Ohta et al. (2011). That increased viscosity improves accuracy of the Fritz bubble size for quasi-static cases might be explained by the nature of viscosity to hamper the capillary force driven dynamics during the neck pinch-off process.

Another observation from the computed results with $Oh$ = 0.2895 indicates that in a liquid of higher viscosity the detached bubbles exhibit little surface deformations from the spherical shape, as expected from viscous suppression of surface-tension driven oscillations. Here the computed bubble rising velocity is in the range of 0.1 to 0.2 m/s, comparable to those computed by Ohta et al. (2011). Thus, with $\rho$ = 1223.8 kg m$^3$, $\sigma$ = 0.066 kg s$^{-2}$, and $\mu$ = 0.126 kg m$^{-1}$ s$^{-1}$ for a bubble of $D$ = 4 mm and rising velocity of 0.2 m/s, the estimated rising bubble Reynolds number would be $Re \sim 9.7$ and Weber number $We \sim 3.7$ which would not cause substantial steady-state bubble surface deformation (cf. Feng 2007), either.

3.3 Air bubbles in a liquid of lower surface tension

So far the computed cases are mostly relevant to air bubbles in aqueous liquids with surface tension around $\sigma \sim 0.07$ kg s$^{-2}$, because most available experimental data are acquired with the air-water systems for laboratory convenience. The case of reduced surface tension is computed here for $D$ = 1 mm and $\theta_0$ = 45° at $U$ = 0.05 m/s with $\sigma$ = 0.035 kg s$^{-2}$ (as representative to many non-aqueous solvents) and everything else the same as that for the air-water system. This is a purely computational benchmark study against the corresponding case for $\sigma$ = 0.07 kg s$^{-2}$ with $\Delta T$ = 274 ms and $V$ = 10.76 mm$^3$ as in table 4. The resulting $\Delta T$ becomes 152.5 ms with $V$ = 5.99 mm$^3$ which is about 10% greater than 0.5 x 10.76 = 5.38 mm$^3$ but slightly closer than the air-water case to that estimated by (1) for the quasi-static situation. Again, the Fritz formulas in (1), based on a static force balance between surface tension and buoyancy, may only be useful for rough estimates rather than accurate predictions in specific applications, because other effects could be more influential especially with reduced surface tension.

## 4 Summary

The OpenFOAM® VOF computational model presented here can yield numerical results in general qualitative agreement with most of those reported by previous authors, which in some sense verifies the model validity. For example, the computed data tabulated in tables 1 to 4 consistently show that the detached bubble size tends to increase with the gas flow rate, orifice size, and liquid contact angle. The case computed in subsection 3.3 indicates that the bubble size also increases with surface tension, qualitatively consistent with the physical expectation suggested by (1). There exists a critical gas flow rate above which detached bubbles will combine via coalescence and exhibit the bubble pairing phenomenon. The detached bubble size seems insensitive to the liquid contact angle when it is smaller than a critical value (referred to as Mode A by Gerlach et al. 2005), but can increase substantially at larger contact angles which result in the contact line movement on the horizontal outside wall of orifice (Mode B per Gerlach et al. 2005). To avoid deeply involving into complexities with the Mode B bubble formation process, the present work mainly focuses on analyzing the Mode A cases with a few Mode B cases computed for brief demonstrative purposes. Operating with the Mode A can be accomplished by using orifice materials that allow adequate liquid wetting, and is desirable for the convenience of process control and minimizing the bubble size.

On the other hand, the present computational study also reveals certain details apparently not yet recognized by previous authors. Majority publications in the literature tend to claim that the detached bubble volume would become about the same as the Fritz value $V_F$ in (1) irrespective of gas flow rate when $Q / Q_{crit} < 1$, whereas the present study shows a linear relationship between $V$ and $Q$ with noticeable slope increasing with orifice size (cf. figure 6 and table 6 based on results for $\theta_0 = 45°$ when the liquid `wets' the orifice walls).

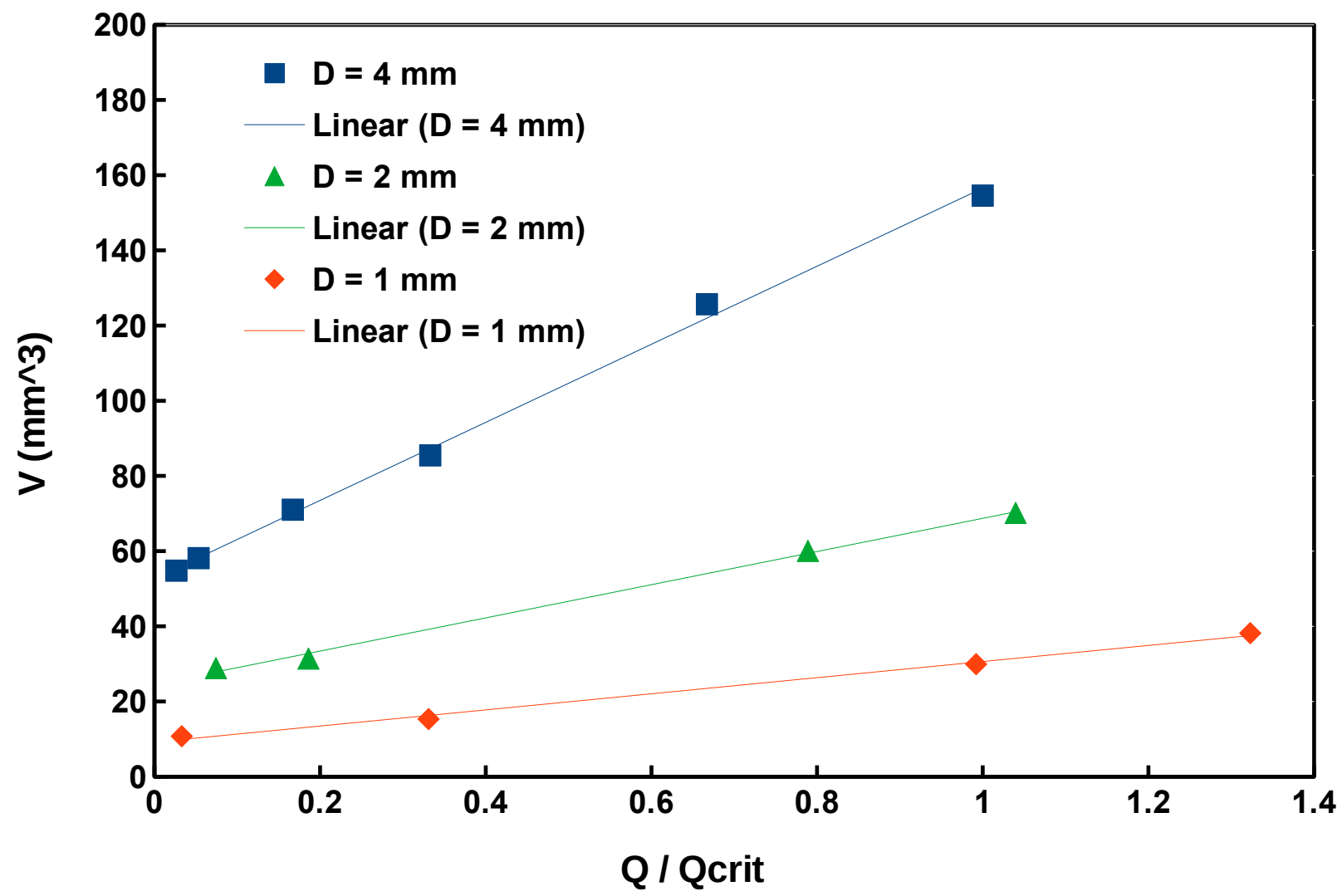

Figure 6. Computed data for $D = 1, 2, 4$ mm with corresponding fitted lines of $R^2 > 0.995$

Table 6. Fitted values of $a$ and $b$ in linear relationship of $V$ versus $x = Q / Q_{crit}$ as $V(x) = a\,x + b$ for $D = 1, 2, 4$ mm with $x$ up to 1 (or not too far above 1), based on results for $\theta_0 = 45°$.

| $D$ (mm) | $Q_{crit}$ (μL/s) | $a$ ($mm^3$) | $b$ ($mm^3$) | $V_F$ ($mm^3$) |
|---|---|---|---|---|
| 1 | 1187.09 | 21.446 | 9.179 | 22.42 |
| 2 | 2115.15 | 44.239 | 24.493 | 44.83 |
| 4 | 3768.77 | 103.84 | 52.710 | 89.67 |

In terms of coefficients $\phi_a$ and $\phi_b$ in (7), we would have $\phi_a$ = 13.52, 5.53, 2.58 and $\phi_b$ = 0.41, 0.55, 0.59, respectively for $D$ = 1, 2, 4 mm. To account for the orifice-size dependence, both $\phi_a$ and $\phi_b$ may be approximated as functions of $D/L_\sigma$ such that

$$\phi_a \simeq 0.415\left(\frac{L_\sigma}{D}\right)^{6/5} \text{ and } \phi_b \simeq 0.54\left(\frac{D}{L_\sigma}\right)^{1/4}, \tag{8}$$

where $(D/L_\sigma)$ turns out to be the square root of the Bond number $Bo=(\rho g D^2/\sigma)$ for measuring the relative importance of gravitational forces compared to the capillary forces. The trends of $\phi_a$ decreasing while $\phi_b$ increasing with $(D/L_\sigma)$ or $Bo$ indicated by (8) would mean that reducing the orifice size $D$ enhances the relative effects of capillary force driven dynamics on detached bubble volume, in view of (7). According to present model findings, the effects of capillary dynamics tend to reduce the bubble size from that of the Fritz value. Such computed bubble volumes for $Q / Q_{crit} << 1$, generally smaller than $V_F$ , seem in good agreement with the experiments of Bari & Robinson (2013) for quasi-static bubble growth with fixed contact line. If further independently confirms, this finding could be a good news because smaller bubbles with a given orifice size is often desirable for heat and mass transfer applications.

However, having smaller bubbles also requires using relatively lower gas flow rates (e.g., with $Q / Q_{crit} < 0.1$), which would compromise the bubble throughput per orifice as an important factor to consider in process development. Also according to the present findings, the critical contact angle $\theta_c$ dividing Mode A and Mode B regimes decreases with reducing the orifice size and with increasing the gas flow rate. This would make the usage of reduced orifice size for generating smaller bubbles more challenging and practically difficult with constant gas flow especially for $D$ < 1 mm. In fact, the experiments of Mohseni et al. (2020) with laser machined sub-millimeter orifices on stainless steel plates in water with measured $\theta_0 \sim 75^o$ demonstrated that bubbles detached from a $D$ = 0.4 mm orifice could have $V \sim$ 21, 27, 32 mm$^3$ at $Q$ = 167, 500, 833 μL/s (with $Q_{crit} \sim$ 553 μL/s), generally larger than that from a $D$ = 1 mm orifice as a consequence of much higher gas flow rate than Q = 39.27 μL/s (in table 4).

Besides gaining further understanding of the complex bubble formation behavior, the usage of open-source CFD software for the present study provides an important merit: it is accessible to everyone who wants to examine and modify the source code. It could open up new opportunities for future investigation of effects due to various numerical algorithms in resolving differences in published model results. Multiple researchers are enabled to conveniently compare notes and collaborate efforts. In this regard, the present results could offer valuable comparison benchmarks for future model refinement and development.

**Statements and Declarations**

**Competing interests** The author has no competing interests to declare.